# Nonlinear photonics of fullerene solutions

E.F.Sheka[1*], B.S.Razbirin[2**], A.N.Starukhin[2], D.K.Nelson[2], M.Yu. Degunov[2], R.N.Lyubovskaya[3], P.A.Troshin[3], N.V.Kamanina[4***]

[1] *Peoples' Friendship University of the Russian Federation, General Physics Department, Laboratory of Computational Nanotechnology, 117198 Moscow, Russia*
[2] *A.F. Ioffe Physical-Technical Institute, RAS, Department of Solid State Optics, 194021 St.Petersburg, Russia*
[3] *Institute of Problems of Chemical Physics, RAS, Laboratory of Synthesis of Multifunctional Organic Compounds, 142432 Chernogolovka, Russia*
[4] *S.I.Vavilov State Optical Institute, 199034 St.-Petersburg, Russia,*

[*] e-mail: sheka@icp.ac.ru
[**] e-mail: b.razbirin@mail.ioffe.ru
[***] e-mail: nvkamanina@hotmail.com

Newly observed enhanced linear optical features of fullerene solutions (Raman scattering and one-photon luminescence) are due to clusterization of fullerene molecules themselves as well as their composites with solvent molecules. A direct connection between the enhanced linear effects and nonlinear behavior of the solutions is discussed and empirical and computational tests of the solutions nonlinear optics efficacy are suggested.

Nanophotonics of fullerenes is mainly associated with the impact of low doping of fullerenes on characteristics of nonlinear optical (NLO) media and, as a result, of NLO devices. Obvious electromagnetic nature of the NLO effects enhancement and a direct influence of electric field on the enhancement witness convincingly the recruitment of charge states in the considered features. These states have been explained by the formation of charge transfer complexes due to donor-acceptor (DA) interaction between fullerenes and host matrices (see extended discussions during one of the first international conference on the matter [1]). Further investigations allowed obtaining a direct confirmation of the complexes formation [2-4] where fullerenes manifest themselves as both donors and acceptors of electron. The latter is a result of a specific DA duality of fullerenes due to their low ionization potentials $I$ and high electron affinities $\varepsilon$.

However, in spite of unquestionable DA character of the fullerene impact on the NLO properties of matrices, a microscopic picture that could 1) explain a non-monotonic dependence of the NLO response on the fullerene concentration; 2) get the answer why the fullerene impact depends on the chemical nature of dopants if modified; 3) suggest a quick testing of the efficiency of the media with fullerene towards the expected NLO effect has retained obscure. Recently observed new spectral events in diluted fullerene solutions, namely, fullerene enhanced Raman scattering (FERS) and solvent enhanced luminescence (SEL) [5] have stimulated us to look for an unique explanation that could unite these events and NLO properties of fullerene-doped matrices. The electromagnetic theory of enhancement (enhancement local field model) [6] that showed its usefulness in describing nonlinear optical effects in matrices consisted of metallic sols may be a basis of such unified consideration.

The electromagnetic theory describes the enhancement in terms of the local fields experienced by polarizable objects which can be much larger than the applied fields. In case of metallic sols, the effect occurs when the field frequency is resonant with those of local surface plasmons. The theory predicts that the enhancement concerns both linear and nonlinear optic effects [6]. These effects are catalogued and conveniently described in the dipole limit by a power series expansion of the induced microscopic dipole moment $p$ over the local field $E_l(\omega)$

$$p = \alpha E_l + \beta E_l E_l + \gamma E_l E_l E_l + ..... \quad (1)$$

The expansion coefficients $\alpha, \beta, \gamma$ are linear, second order, and third order generalized polarizabilities. As well known, the quantities are responsible for linear refractive index, linear absorption and luminescence ($\alpha$, spontaneous Raman scattering is caused by a periodic change in $\alpha$ induced by the vibration-frequency modulation of the applied fields in the form $\alpha = \alpha_0 + (\partial \alpha / \partial Q)Q$, where Q is one of the object vibration coordinate), the second harmonic generation, sumfrequency mixing, and optical rectification of different types ($\beta$), and the third harmonic generation, various four-wave mixing processes, optically induced changes to the linear refractive index, and two-photon absorption and luminescence ($\gamma$). All induced dipoles collectively interact and produce a macroscopic polarization $P$ that is given by

$$P(\omega) = L'(\omega)\chi^{(1)}L(\omega)E(\omega) + L'(\omega)\chi^{(2)}L(\omega_1)L(\omega_2)E(\omega_1)E(\omega_2)$$
$$+ L'(\omega)\chi^{(3)}L(\omega_1)L(\omega_2)L(\omega_3)E(\omega_1)E(\omega_2)E(\omega_3) + ... \quad (2)$$

Local field factors $L'$ and $L$ are introduced to describe the amplification of the incident and outgoing fields, respectively. The macroscopic generalized susceptibility $\chi^{(i)}$ contains the sum over all microscopic generalized polarizabilities in the interaction volume. Therefore, according to the electromagnetic model, local field factors $L'$ and $L$ are connecting links between linear and nonlinear optic effects of the polarizable object. So that the study of enhanced linear effects throws light into a microscopic origin of nonlinear effects and vice versa.

Local field factors depend sensitively upon the size and shape of the polarizable object, the orientation and location of the molecules that form the object, the effect of neighboring structures and many other structural features. A correct determination of the factors is hardly possible. However, some evaluation of the value scale can be taken from the estimation of the factor for an isolated sphere of diameter much less than the wavelength [7]

$$L(\omega) = 3/[\varepsilon(\omega) + 2\varepsilon_m] \quad (3)$$

where $\varepsilon(\omega)$ and $\varepsilon_m$ are real parts of the complex dielectric constants of the polarizable object and surrounding, respectively. Thus $L(\omega)$ becomes large for $\varepsilon(\omega) = -2\varepsilon_m \approx -2$. This is the resonant condition for plasma oscillation of the sphere. The condition is approximately valid for the polarizable object of an arbitrary shape.

To apply the enhanced local field model to the characterization of fullerene solutions we have to determine what the polarizable object is in this case, what is responsible for the excitation of the electron-hole plasma, and how is the interval of the plasma resonant frequencies as well as to estimate the resonant conditions for the local field factors.

As for the polarizable object, recently observed spectral features [5, 8] have shown that clusters of fullerenes provide the entity. It should be noted that the solute molecule clusterization is rather typical for diluted fullerene solutions [9-11]. Its efficacy is governed by many microscopic and macroscopic factors. Among the latter there are temperature, viscosity and internal structure of the solvent, particularly in the case of substances that are susceptible to mesophase formation. At the microscopic level, the cluster formation is governed by the energy of intermolecular interaction or coupling energy $E_{cpl}$ of a pair of molecules, both solute molecules themselves (*sol-sol*) and solute and solvent molecules (*sol-solv*). Table 1 contains a list of quantum chemically calculated $E_{cpl}$ values of a set of *sol-sol* and *sol-solv* pairs involving different $C_{60}$ derivatives and toluene as a solvent studied in the paper. Since DA interaction dominates in the total intermolecular interaction, both the formed pairs and enlarged clusters are charge transfer complexes. Their photoexcitation is followed by the electron transfer within the cluster from one molecule to another thus producing a space confinement for charge transfer excitons (CTEs). Therefore, *sol-sol* and *sol-solv* fullerene clusters can be considered similarly to colloidal metallic particles where local plasmon excitation is substituted by the formation of local CTEs (LCTEs). Due to weak intermolecular interaction, the LCTEs energy spectrum of both $C_{60}$- and $C_{70}$-based clusters should not be too far from the spectra of CTEs of $C_{60}$ and $C_{70}$ crystals that are located in the visible region [12, 13], which perfectly suits the convenient laser pumping.

Equilibrium configurations of some clusters of $C_{60}$-based fullerenes listed in Table 1 are shown in Fig. 1. Assuming a spherical envelope of the structures, we can use Eq.3 to estimate the local field factor. The dielectric losses study of $C_{60}$ crystal revealed a rather broad energy interval in the visible region where $\varepsilon(\omega)$ changes from -5 to -2 [14] thus fitting the resonant conditions with respect to pumping lasers. There is a good reason to assume that the resonant conditions are maintained for the majority of $C_{60}$-based fulleroids. Therefore, both the location of the LCTE energy spectrum and local field factor resonance in the visible region makes the fullerene-doped matrices to serve as perfect objects for enhancing optical effects.

As said before, within the enhanced local field model, the nonlinear and linear optical effects are closely linked. This stimulated us to look at peculiarities of linear optic properties of fullerene solutions from the viewpoint of their feasible exploitation to predict the NLO properties of the fullerene-based materials. Enhanced luminescence and enhanced Raman scattering were the main targets of our investigations. The first success has been achieved when studying spectral properties of fullerenes **I**, **II**, and **III** in toluene solutions [5]. Fig. 2 shows the emission spectra of the three solutions which differ in the efficacy of the *sol-sol* clusterization. The latter increases when going from solution **I** to **III** (see Table 1). The red spectra of the solute molecules in the figure are taken as inner standard to scale the emission spectra. The first common feature of the spectra is the presence of a "blue" component ("blue spectrum") located in the visible region, whose appearance depends on the chemical composition of the solution. The spectrum is characteristic for the solution only since the solute and solvent molecules emit light in the red-IR and UV regions, respectively. The second common feature is related to the dependence of the blue spectrum on the excitation wavelength $\lambda_{exc}$. And the third one concerns an evident dependence of the blue spectrum intensity on the efficacy of the fullerene clusterization. When the clusterization does not occur (solution **I**), the blue spectrum is very weak and presents a spontaneous Raman scattering from toluene. When the clusterization is appreciable, though is not too significant (solution **II**), the Raman scattering from toluene takes features of the enhanced spectrum. When the clusterization is well pronounced (solution **III**), the blue spectrum presents highly intense enhanced Raman scattering from the fullerene molecules integrated into the clusters and is attributed to FERS [5]. Common to solutions **II** and **III** is that the blue spectrum intensity increases when $\lambda_{exc}$ changes from 514.5 nm to 476.5 nm. We believe that this effect is due to moving $\lambda_{exc}$ under this change into the depth of the LCTE density of states (DOS), thus strengthening the resonant conditions, of *sol-sol* clusters that is presented in Fig.2 by the CTE DOS of $C_{60}$ crystal.

Additional confirmation of a particular role of the solute molecules clusterization in linear spectral effects is presented in Fig. 3. The figure shows emission spectra of the three solutions under the excitation of high level states of both solute and solvent. Previously observed blue spectra are absent under this condition and one observes the red spectrum of the solute molecules and near-UV spectrum of the solvent. The emission spectrum of toluene (in fact, it is due to small irremovable admixture of benzaldehyde [15]) occupies practically the same position as the LCTEs energy spectrum of *sol-sol* clusters, thus meeting the relevant resonant requirements [6]. As a result it demonstrates a considerable enhancement and is transformed to SEL when going from solution **I** to solution **III** in parallel to increasing efficacy of the fullerene clusterization.

The peculiarities of linear optic events in fullerene-doped matrices go beyond the observed FERS and SEL. Thus, if the formation of *sol-sol* clusters dominates in solutions **II** and **III**, fullerene **IV,** as seen from Table 1, forms both *sol-sol* and *sol-solv* clusters (see Fig.1). These results in a new manifestation of the blue spectrum of solution **IV** as it is shown in Fig. 4 (see detailed discussion of the spectra in [8]). As in the case of solution **III**, solution **IV** is characterized by intense blue spectrum in the region of 15000-20000 cm$^{-1}$. However, oppositely to **III,** the blue spectrum of **IV** consists of another pair of components, namely, of enhanced Raman scattering from toluene, as previously, and enhanced broad-band luminescence. The spectrum disappears when toluene is substituted by carbon tetrachloride (see dotted plotting in the figure). As previously, the excitation of both components of the spectrum is in resonance with the LCTEs states of the *sol-sol* type. This explains a quasi-resonant growing of the intensities of both components when $\lambda_{exc}$ decreases. The LCTEs states of the *sol-solv* type is red-shifted due to lowering the toluene ionization potential with respect to that of fullerene (see light grey blocks in Fig.4). This shift is enough to satisfy the resonant conditions for the *sol-sol* cluster luminescence that is situated over *sol-solv* LCTEs states, which results in fullerene cluster enhanced luminescence (FEL).

Therefore, the appearance of intense blue emission spectrum in a fullerene-doped matrix either in the form of SERS and SEL of solvent or in the form of FERS and FEL of fullerenes points to the efficacy of the fullerene clusterization that provides a considerable amplification of the incident and outgoing light. Obviously, the blue spectrum appearance depends on many factors, among which there are chemical composition of clusters formed, cluster size and their distribution over the matrix, surrounding temperature that greatly influences the clusterization process and causes upward and/or downward shift of the energy spectrum of LCTEs, etc. However, disclosing the blue spectrum can serve as a convincing test on the applicability of the studied matrix to nonlinear optics. Therewith, the spectrum intensity undoubtedly points to the efficacy of the expected NLO applications.

To check the conclusion, we have tested $C_{70}$-doped cyanobiphenyl ($C_{70}$-*cbph*) solution. Oppositely to toluene, *cbph* itself is widely used as either an effective limiter or display and reversible diffractive NLO element [16]. Similarly to fullerenes, *cbph* may act as both donor and acceptor of electron whilst weaker than fullerenes. Nevertheless, as shown by the performed QCh calculations, DA interaction governs the formation of a molecular *solv-solv* pair with $E_{cpl}$ = -0.83 kcal/mol. Therefore, the DA clusterization of the solution is rather effective and one might expect the appearance of the characteristic emission spectrum in the visible region additionally to the intrinsic UV luminescence of *cbph* molecules when the corresponding resonant conditions are satisfied. Actually, this spectrum (identical to shown in Fig.5a) is observed when $\lambda_{exc}$ changes from 514.5 to 476.5 nm alongside with the enhanced Raman scattering from *cbph* and is substituted by the molecular UV spectrum at $\lambda_{exc}$ = 337.1 nm. It shows all features characteristic for FEL but in the absence of fullerenes should be attributed to the solvent enhanced cluster luminescence (SECL) as caused by the *solv-solv* clusterization. Doping with fullerene $C_{70}$ introduces a new partner in a complicated scheme of the DA interactions with $E_{cpl}$ of -0.03 kcal/mol for *sol-solv* pairs and a positive $E_{cpl}$ for *sol-sol* pairs. Due to small coupling energy, the *sol-solv* clustering manifests itself via an additional blue spectrum only at low temperature (LT) (Fig.5b) while at room temperature the clusters are dynamically unstable and only *solv-solv* clusters are spectrally observed. LT blue spectrum of the solution possesses all main features similar to those of solution **IV** discussed earlier. It consists of two components, one of which is the solvent Raman scattering spectrum while the other exhibits a luminescence behavior. However, oppositely to a coordinated following of the two component intensity to the change in $\lambda_{exc}$, in the case of solution **IV**, a non-coordinated behavior is observed for the $C_{70}$-*cbph* solution. As seen in Fig.5b, at fixed intensity of the red $C_{70}$ molecular spectrum, the Raman scattering intensity increases while the blue luminescence intensity decreases when $\lambda_{exc}$ changes from 514.5 to 476.5 nm. This points to the fact that polarizable nanoobjects responsible for the enhancement of Raman scattering and blue luminescence are of different origin with different block-like spectra of LCTEs states. The latter may lie either below or above a fixed range of laser pumping frequencies which provides a different direction of the $\lambda_{exc}$ movement towards the resonance.

As a whole, a "blue-spectrum" test of the solution is obviously positive that well correlates with a large use of both *bphn* itself and $C_{70}$-*cbph* composite in NLO applications. Additionally, chemically obvious, a low solubility of fullerenes in a particular solvent may be considered as another empirical test of effective clusterization of fullerenes.

Besides the two empirical tests, a computational one can be suggested. That concerns quantum-chemical calculations of the coupling energy $E_{cpl}$ in pairs of *sol-sol, sol-solv*, and *solv-solv* types. Evidently, the test is the most effective if $E_{cpl}$ is the main governing factor of clusterization as it was in toluene solutions. As follows from Table 1, $E_{cpl}$ values that exceed 0.5 kcal/mol by absolute value, make one to speak about an efficient fullerene clusterization in solution **III** and **IV** and, thus, about a high efficacy of using these systems in NLO applications. This is definitely a limit case of strong interaction when weak effects of macroscopic factors can be ignored. However, when the interaction is weak, as is the case of the $C_{70}$-*cbph* matrix, the macroscopic factors can be rather important, particularly for mesomorphic matrices with easily changed molecular arrangement.

The approach presented above can be spread over solutions with carbon nanotubes (CNTs). As known empirically [17], small additions of CNTs cause the enhancement of NLO properties of active matrices analogously to fullerenes. Among other possible reasons, DA interaction between CNTs themselves and solvent molecules can be responsible for the effect. Actually, CNTs are both good donors and acceptors of electrons with DA characteristics close to those of fullerenes. This provides a possibility of the formation of CNTs associates that possess properties of charge transfer complexes, on the one hand, and are characterized by LCTEs states with the energy spectrum in the near IR region, on the other. The local field factors of formed associates should be close to those of metallic nanorods and/or nanorices that are good amplifiers of light [18]. Enhanced linear spectral effects of CNTs solutions have been under question so far.

The work is financially supported by the RFBR (grant № 07-03-00755 and partially grant № 08-02-00966).

**Table 1.** Ionization potentials and electron affinities of fullerenes and coupling energy of fullerene clusters

| Fullerenes[1] | $I$, eV | $\varepsilon$, eV | $E_{cpl}\{(X)_n\}$ of *sol-sol* clusters[2], kcal/mol | | | | |
|---|---|---|---|---|---|---|---|
| | | | 2[3] | 3 | 4 | 5 | 6 |
| **I** 1-methyl-2(4-pyridine)-3,4-$C_{60}$fulleropyrrolidine | 9.68 | 2.48 | 2.21; 1.16; 1.06 | - | - | - | - |
| **II** $C_{60}$ fullerene | 9.87 | 2.66 | -0.52 | - | - | - | -2.74 |
| **III** azyridine [$C_{60}$] fullerene | 9.79 | 2.57 | -0.09; -1.26 | -2.73 | -4.32 | -7.94 | -8.42 |
| **IV** ethyl ester of the [$C_{60}$]fullerene acetic acid | 9.84 | 2.59 | -3.66 | -8.01 | -3.21 | - | - |

| | | | $E_{cpl}\{(X)_n(T)_m\}$ of *sol-solv* clusters[4], kcal/mol | | | | |
|---|---|---|---|---|---|---|---|
| | | | $(IV)_1(T)_1$ | $(IV)_2(T)_1$ | $(IV)_2(T)_2$ | $(IV)_3(T)_2$ | $(IV)_4(T)_4$ |
| **T** | 9.34 | -0.56 | -0.64 | -4.23 | -6.02 | -9.25 | -5.81 |

[1] Equilibrium structures of fullerenes **I**, **II**, **III**, and **IV** are shown as inserts in Figs. 2 and 4
[2] QCh calculations have been performed in semiempirical approach by using AM1 version of the CLUSTER-Z1 software (see details in [8]). The data for ionization potentials are traditionally somewhat overestimated while electron affinities well suit experimental data. The coupling energy $E_{cpl}\{(X)_n\}$ of $(X)_n$ cluster was determined following the relationship $E_{cpl}\{(X)_n\} = \Delta H\{(X)_n\} - n\Delta H\{X\}$ where $\Delta H\{(X)_n\}$ and $\Delta H\{X\}$ are heats of formation of $(X)_n$ cluster and $X$ fullerene (X= **I, II, III, IV**), respectively.
[3] Data present the largest and smallest energy coupling values that depend on mutual disposition of molecules in a pair.
[4] $E_{cpl}\{(X)_n(T)_m\} = \Delta H\{(X)_n(T)_m\} - n\Delta H\{X\} - m\Delta H\{T\}$. Here $T$ marks solvent molecules, toluene in the case.

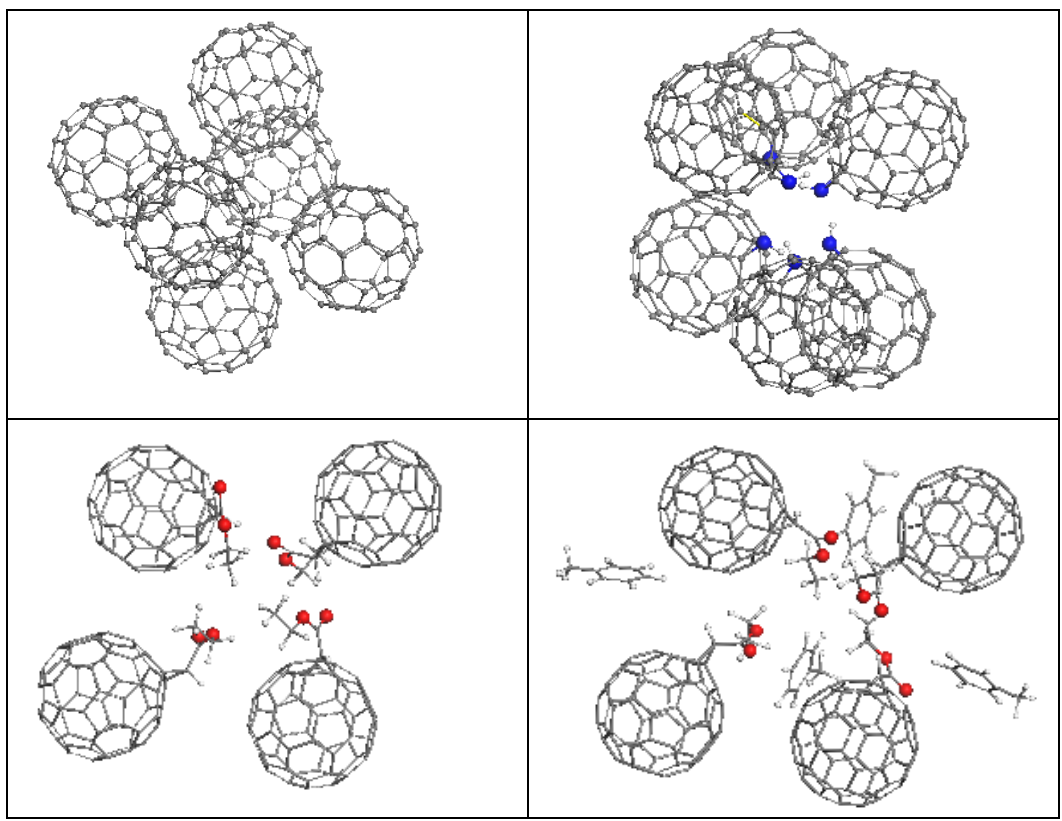

**Figure 1.** Equilibrium structures of clusters (**II**)$_6$ and (**III**)$_6$ (top) and (**IV**)$_4$ and (**IV**)$_4$ (**T**)$_4$ (bottom). The structures were obtained in due course of total optimization of the cluster configurations when seeking the energy minimum. Big blue, red and small white spheres mark nitrogen, oxygen, and hydrogen atoms of the addends, respectively.

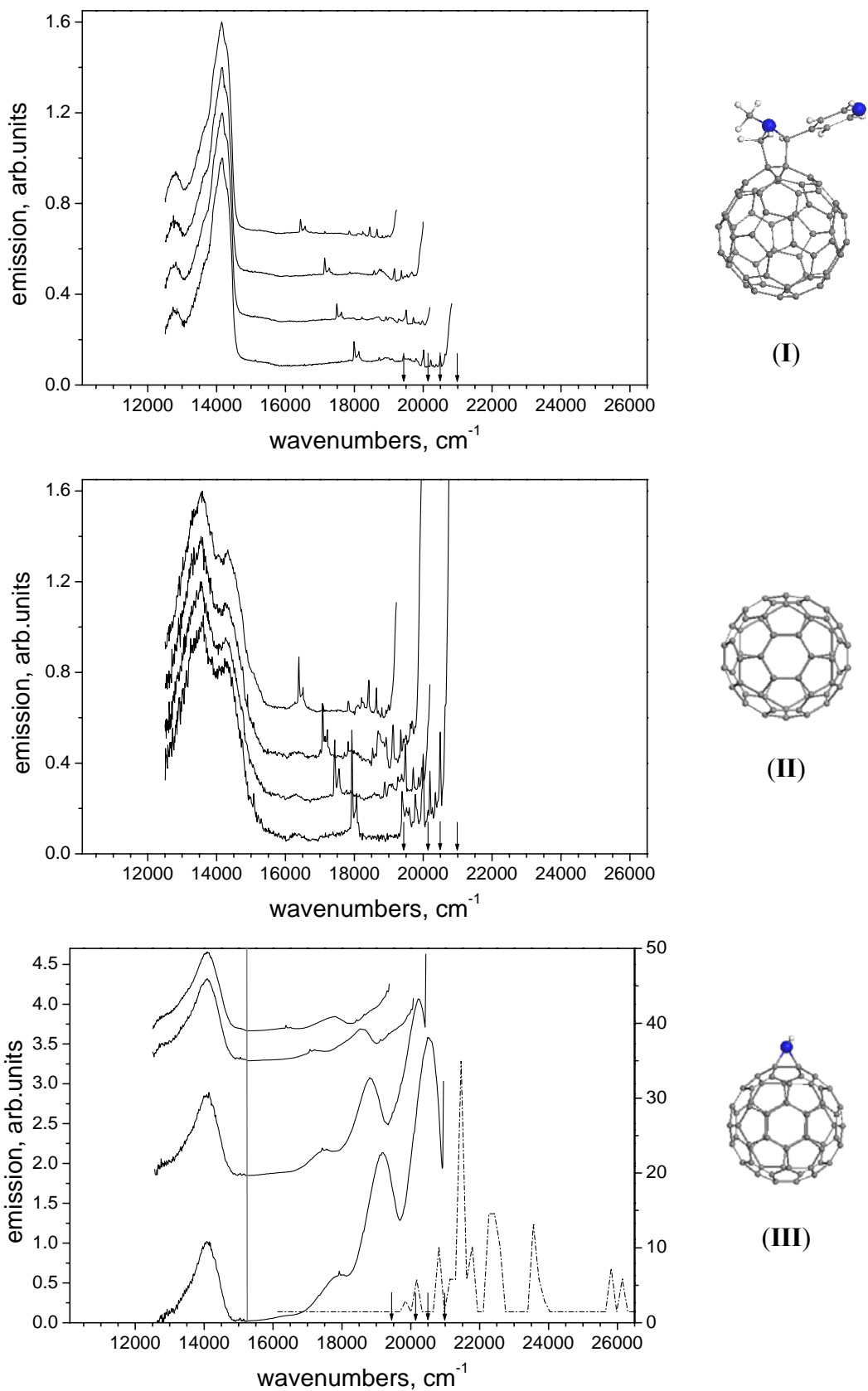

**Figure 2.** Emission spectra of fullerenes **I**, **II**, and **III** in toluene at 80K. At each panel, spectra from bottom to top correspond to $\lambda_{ex}$ 476.5, 488.0, 496.5, and 514.5nm, respectively, marked by arrows from right to left. Dotted curve plots the density of CTE states of $C_{60}$ crystal courtesy by Dr.A.Eilmes and Dr.B.Pack, Jagiellonian University, Crakow.

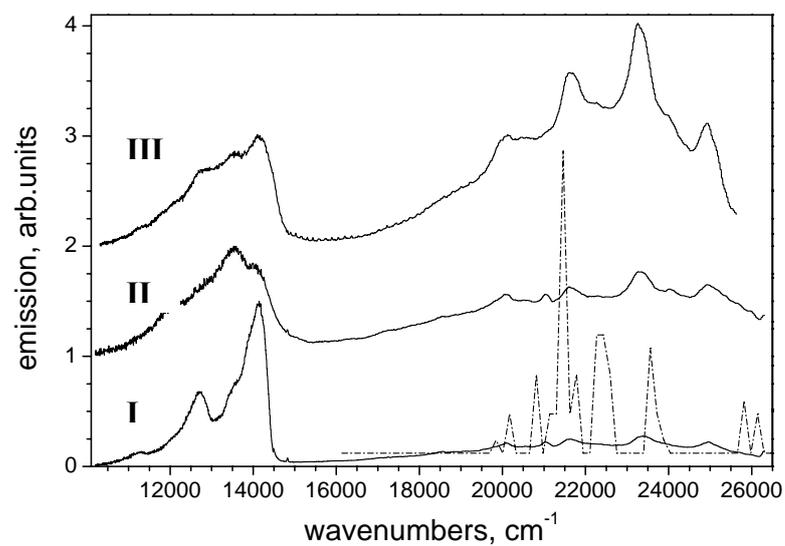

**Figure 3.** Emission spectra of fullerenes **I**, **II**, and **III** in toluene at 80K, $\lambda_{ex}$ = 337.1nm. Dotted curve plots the density of CTE states of $C_{60}$ crystal courtesy by Dr.A.Eilmes and Dr.B.Pack, Jagiellonian University, Crakow.

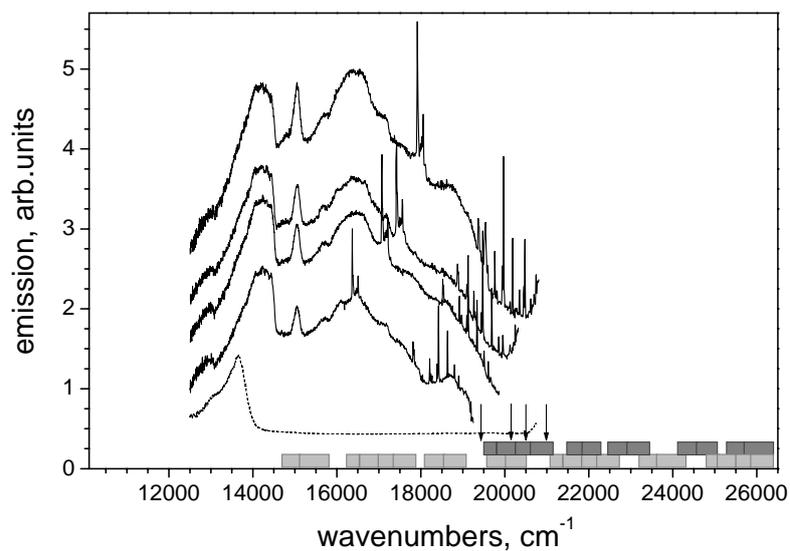

**Figure 4.** Emission spectra of fullerene **IV** in toluene at 80K. Solid curves from bottom to top correspond to $\lambda_{ex}$ 514.5, 496.5, 488.0, and 476.5 nm, respectively, marked by arrows from left to right. Dotted curve plots emission spectrum of fullerene **IV** in carbon tetrachloride at 80K and $\lambda_{ex}$ = 476.5 nm. Dark and light grey boxes schematically depict the energy spectra of the LCTEs of the *sol-sol* and *sol-solv* types, respectively.

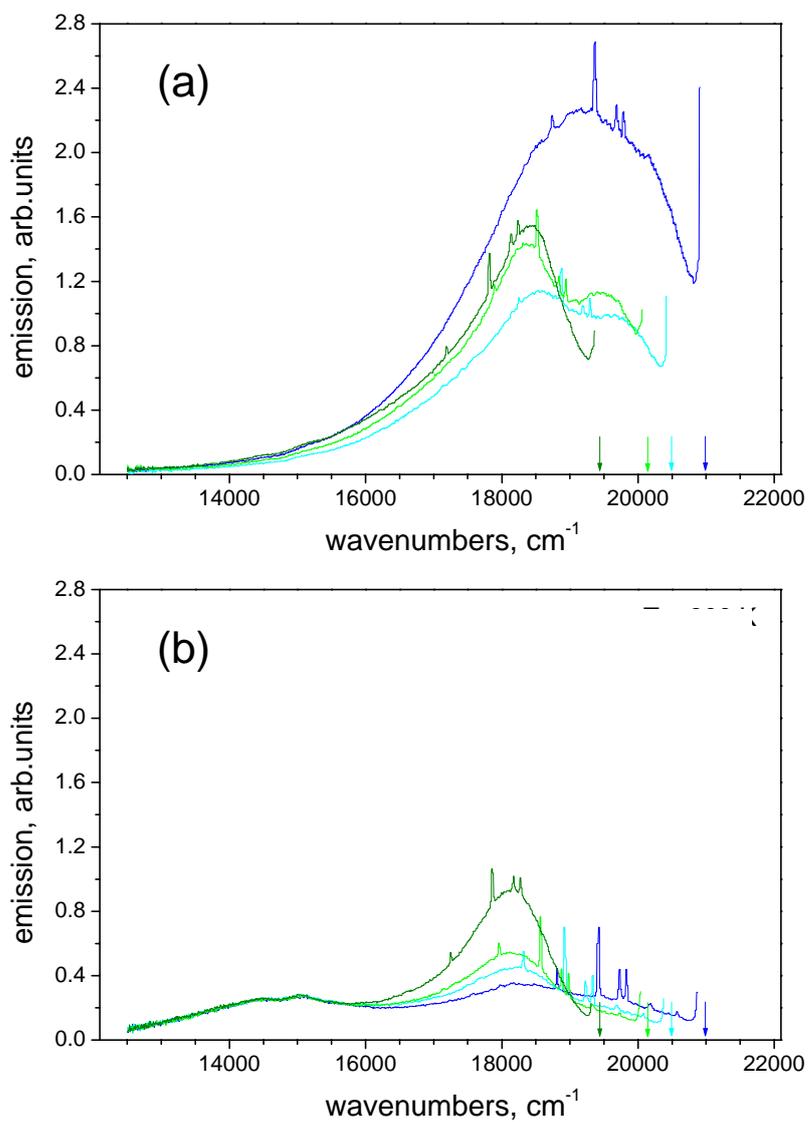

**Figure 5.** Emission spectra of the $C_{70}$-doped cyanobiphenyl. Spectra from top to bottom in (a) and from bottom to top in (b) correspond to $\lambda_{ex}$ 514.5, 496.5, 488.0, and 476.5 nm, respectively, marked by arrows from left to right. a. Isotropic liquid, T=300K. b. Frozen drop at T=80K.